\documentclass{sf2a-conf2025}
\usepackage{graphicx}
\usepackage{hyperref}
\usepackage[]{natbib}  
\usepackage{epstopdf}
\usepackage{subcaption}
\usepackage{amsmath}
\usepackage{cleveref}
\usepackage[rightcaption]{sidecap}

\def\BibTeX{{\rm B\kern-.05em{\sc i\kern-.025em b}\kern-.08em
    T\kern-.1667em\lower.7ex\hbox{E}\kern-.125emX}}
\bibpunct{(}{)}{;}{a}{}{,}  

\defcitealias{Barrault2025ConstrainingDips}{B+25a}
\defcitealias{Barrault2025ExploringField}{B+25b}
\defcitealias{Tokuno2022AsteroseismologyOscillations}{TT22}

\begin{document}

\TitreGlobal{SF2A 2025}


\title{The inertial dip as a window on the convective core dynamics}

\runningtitle{The inertial dip as a window on the convective core dynamics}

\author{L. Barrault}\address{Institute of Science and Technology Austria (ISTA), Am Campus 1, 3400 Klosterneuburg, Austria}

\author{L. Bugnet$^1$}


\author{S. Mathis}\address{Université Paris-Saclay, Université Paris Cité, CEA, CNRS, AIM, 91191, Gif-sur-Yvette, France} 
\author{J.\,S.\,G. Mombarg$^{2}$}

\setcounter{page}{237}


\maketitle


\begin{abstract}
$\gamma$ Dor stars are ideal targets for studies of the innermost dynamical properties of stars, due to their rich
frequency spectrum of gravito-inertial modes propagating in the radiative envelope. Recent studies found that these modes could couple at the core-to-envelope interface with pure inertial modes in their sub-inertial regime, forming the so-called inertial dip in the
period-spacing pattern of these stars. The inertial dip, as formed by core modes, stands out as a unique probe of core properties. We aim in this work to explore the effect of core magnetism on its structure, property of key relevance in modern stellar physics. We describe the outlines of our model and the geometry of the considered field. We give the coupling equation and the variation of the dip shape and location with increasing magnetic contrast between the core and the envelope. We compare our findings to the ones obtained in a hydrodynamical, differentially-rotating case. We show hints at potentially lifting the degeneracy between the signatures of core-to-envelope differential rotation and core magnetic fields. Together, these two cases can be considered as an exploration of different magnetic regimes potentially reached in the core of $\gamma$ Dor stars.

\end{abstract}

\begin{keywords}
asteroseismology – methods: analytical – stars: oscillations – stars: magnetic field – stars: rotation –
convection
\end{keywords}


\section{Introduction}
The fine quality of asteroseismic data provided by the \textit{Kepler} space mission \citep{Borucki2010KeplerResults} has allowed the revolution of magneto-asteroseismology: the detection of magnetic fields from the alteration of oscillation frequencies or amplitudes. $\gamma$-Dor stars are ideal targets for such an analysis, because of their very rich asteroseismic spectrum displaying a variety of radial orders. Their envelope gravito-inertial modes are now proven to bear an influence of envelope magnetism on their frequencies \citep{Dhouib2022DetectingField,Lignieres2024PerturbativeModes, Rui2024AsteroseismicStars}.
\citet{Ouazzani2020FirstRevealed} established that envelope gravito-inertial modes of $\gamma$-Dor stars can couple through the core boundary to core pure inertial modes, in the sub-inertial regime. This interaction causes a characteristic inertial dip in the period-spacing pattern of such stars, i.e. the difference of periods between two modes of same angular structure but successive radial orders, as a function of the period of the modes. The influence of core density stratification, rotation rate, differential rotation have been derived \citep{Saio2021RotationModes, Tokuno2022AsteroseismologyOscillations, Galoy2024PropertiesStars, Barrault2025ConstrainingDips} both numerically and analytically. Gravito-inertial modes could then be used not only to probe radiative envelope properties but also those of the convective core.\\
We aim in this work to present new results of \citet{Barrault2025ExploringField} (hereafter \citetalias{Barrault2025ExploringField}). This work is a generalisation of the theoretical formalism developed by \citet{Tokuno2022AsteroseismologyOscillations} (hereafter \citetalias{Tokuno2022AsteroseismologyOscillations}) in the case of a solid-body, non-magnetic star, later adapted to a bi-layer differential rotation by \citet{Barrault2025ConstrainingDips} (hereafter \citetalias{Barrault2025ConstrainingDips}), to a magnetic configuration, in which we considered a purely toroidal field with a bi-layer Alfvén frequency. A potential future detection of core magnetism on the main sequence would be decisive in understanding the evolution of magnetic fields along stellar ageing, a key unknown of modern stellar physics.


\section{Findings of the analytical study}
\subsection{Outline of the model}

\begin{figure}
\begin{subfigure}{.5\textwidth}
  \centering
  \includegraphics[trim = 0.3cm 4cm 0.3cm 4cm,clip,width=0.6\linewidth]{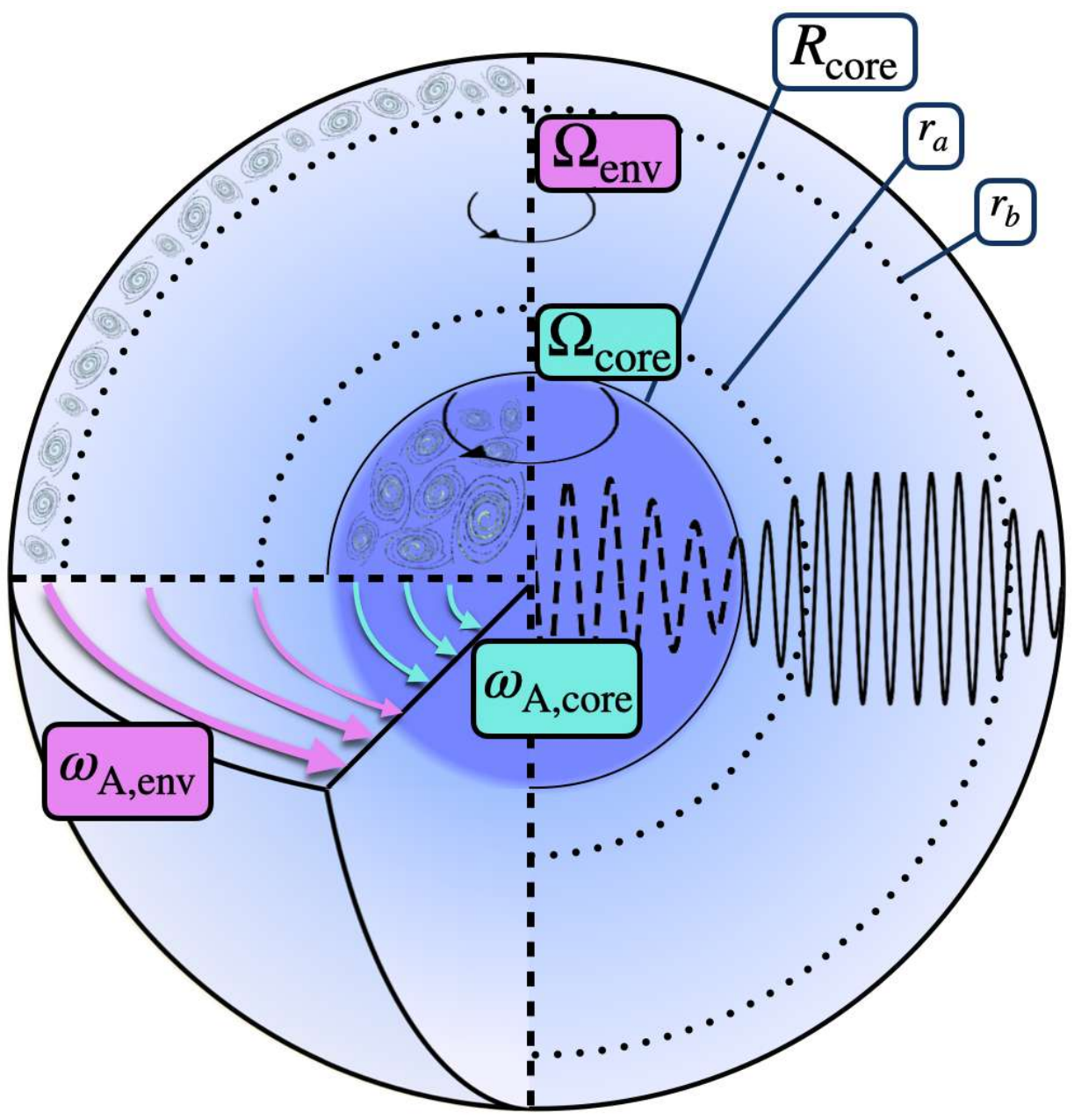}
  \label{fig:disc_Alfv}
\end{subfigure}%
\begin{subfigure}{.5\textwidth}
  \centering
  \includegraphics[trim = 0.2cm 0cm 0.2cm 0cm,clip, width=.9\linewidth]{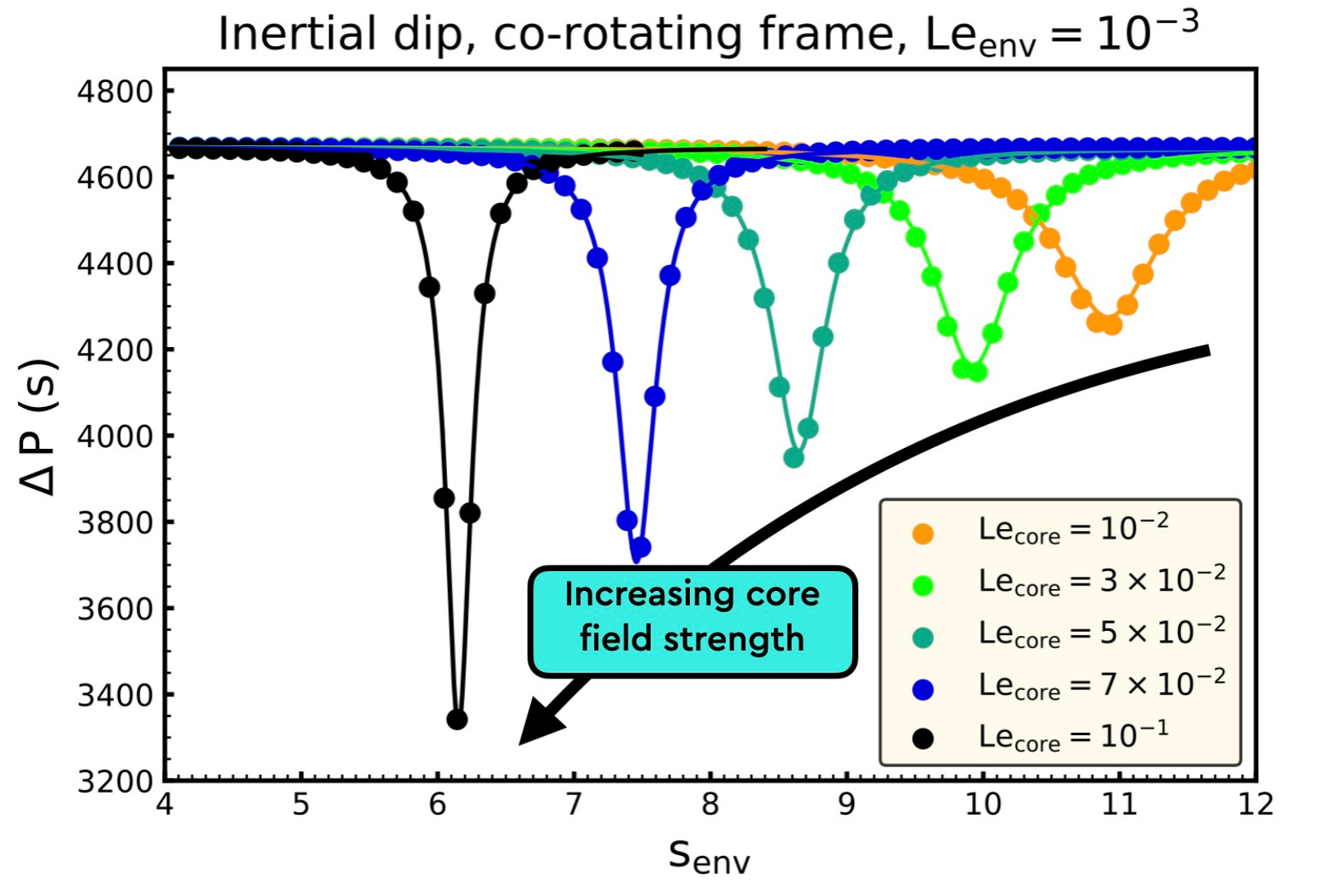}
  \label{fig:sfig2}
\end{subfigure}
\caption{\textbf{Left}: Magnetic star with a bi-layer Alfvén frequency, $\omega_{\rm A, core}$ in the core, $\omega_{\rm A, env}$ in the envelope, and a bi-layer rotation rate, $\Omega_{\rm core}$ in the core, $\Omega_{\rm env}$ in the envelope. The cavity for $m\text{-}g\text{-}i$ modes lies between $r_{\rm a}$ and $r_{\rm b}$ in the radiative zone. They become evanescent in the region $[R_{\rm core};r_{\rm a}]$ when the TARM is applied. $m\text{-}i$ modes propagate in the convective core below the location $R_{\rm core}$. \textit{Extracted from \citetalias{Barrault2025ExploringField}}. \textbf{Right}: Inertial dips overplotted for different core Lehnert number for a fixed envelope Lehnert number $\rm Le_{env} = 10^{-3}$, in a uniformly rotating star, in the frame co-rotating with the envelope. The dots are obtained by solving the coupling equation numerically, and the continuous line by applying a dip profile given in \citetalias{Barrault2025ExploringField}. \textit{Adapted from \citetalias{Barrault2025ExploringField}}.}
\label{fig:1}
\end{figure}

We choose an azimuthal axi-symmetric magnetic field (see left panel of Fig.~\ref{fig:1}) with a bi-layer Alfv\'en angular frequency $\omega_{\rm A}$, such that $\omega_{\rm A} = \omega_{\rm A, core}$ in the convective core and $\omega_{\rm A} = \omega_{\rm A, env}$  in the radiative envelope. The rotation profile is as well bi-layered, with a rotation rate $\Omega_{\rm core}$ in the core and $\Omega_{\rm env}$ in the envelope. This framework allows us to understand the respective effects of rotation or magnetism on the dip formation. This configuration retains a toroidal component present in both the scenarios able to shape the magnetic field in a radiative envelope: a relaxed fossil field \citep[e.g.][]{Braithwaite2006StableInteriors,Kaufman2022TheFields,Becerra2022StabilityStars}, and a Tayler-Spruit like mechanism \citep[e.g.][]{Spruit2002DynamoInterior,Petitdemange2023Spin-downLayers}. We allow for different magnetic field amplitudes from both sides of the boundary, as different magnetic field formation mechanisms are at play in the two zones and can lead to significantly dissimilar amplitudes \citep{Featherstone2009EffectsStars}.\\
As in \citetalias[][]{Barrault2025ConstrainingDips}, we place ourselves in the sub-inertial regime for which the local angular wave frequencies in the two zones are inferior to the inertial frequencies $(2\Omega_{\rm core},2\Omega_{\rm env})$. Magneto-gravito-inertial ($m\text{-}g\text{-}i$) modes can propagate in the core as magneto-inertial ($m\text{-}i$) modes in this framework. We adopt an hypothesis of anelasticity in both the convective core and the radiative envelope, due to the low frequencies of the considered modes. We use as well the Cowling approximation \citep{Cowling1941TheStars}, as the series of prograde dipole modes found in $\gamma$ Dors have a high radial order \citep{Li2020Gravity-modeKepler}.
We assume adiabaticity in the whole region of propagation of both types of modes. We thus neglect heat, Ohmic, and viscous diffusions. We neglect the centrifugal force, as the rotation rate of the innermost layers of the radiative zone is negligible compared to the critical rotation rate \citep{Dhouib2021ThePlanets}. We neglect the indirect effect of the magnetic field on the hydrostatic equilibrium, as $\omega_{\rm A, zone} <\Omega_{\rm zone} \ll N$, $N$ being the Brunt-V\"ais\"al\"a frequency.\\
We adopt the traditional approximation of rotation and magnetism (TARM) in the radiative envelope \citep{Mathis2011Low-frequencyField,Dhouib2022DetectingField}, under the hypothesis of strong envelope stratification.
Under the TARM, $m\text{-}g\text{-}i$ modes propagate between an inner ($r_{a}$) and an outer ($r_{b}$) radial coordinate (See left panel of Fig.\ref{fig:1}). We write the Lagrangian pressure perturbation and the displacement at $r_{a}$ in the JWKB approximation, provided that $N\gg \sigma_{\rm env}$ everywhere in the $m\text{-}g\text{-}i$ mode cavity except at the lower turning point, with $\sigma_{\rm env}$ the local angular wave frequency. We use this expansion at the core boundary limit $R_{\rm core}$, provided that the $N$ gradient near the core is very high, ensuring that $r_{\rm a}$ is close to $R_{\rm core}$. In the convective core, an hypothesis of uniform mean density allows us to isolate $m\text{-}i$  modes in the form of the Bryan-like solutions \citep{Bryan1889TheEllipticity,Malkus1967HydromagneticWaves}. We derive the same quantities as from the envelope side and further match them at $R_{\rm core}$, at a fixed frequency in the inertial frame.
\subsection{Variation of the dip profile with magnetism}

\begin{figure}
\begin{subfigure}{.5\textwidth}
  \centering
  \includegraphics[trim = 0.2cm 0cm 1.2cm 0cm,clip, width=.8\linewidth]{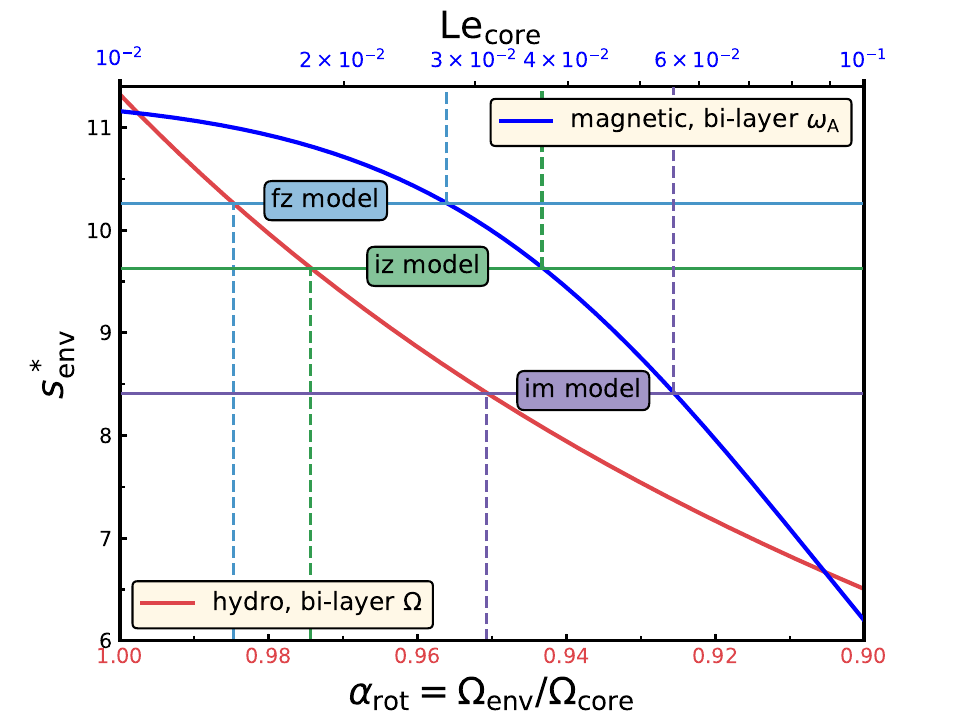}
  \label{fig:disc_Alfv}
\end{subfigure}%
\begin{subfigure}{.5\textwidth}
  \centering
  \includegraphics[trim = 0.2cm 0cm 0.2cm 0cm,clip, width=0.95\linewidth]{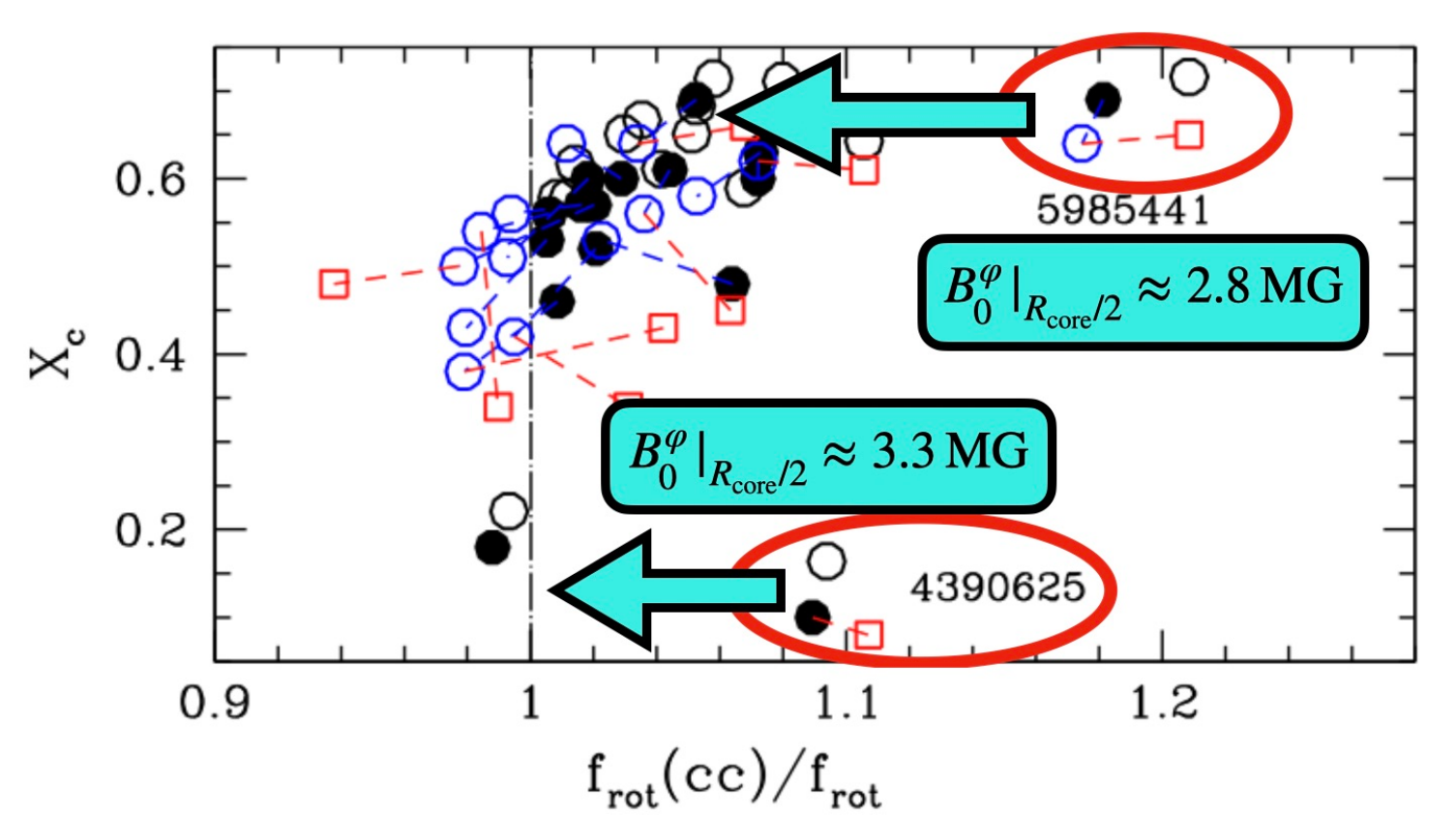}
  \label{fig:Saio}
\end{subfigure}
\caption{\textbf{Left}: Red: Spin parameter of the pure inertial mode in the frame co-rotating with the envelope $s^{*}_{\rm env}$ in the hydrodynamic, differentially-rotating case as function of the amount of differential rotation \citepalias{Barrault2025ConstrainingDips} Blue: the magnetic, bi-layer Alfvén frequency case as function of $\rm Le_{core}$ \citepalias{Barrault2025ExploringField}. Horizontal solid lines correspond to the shift of the dip introduced by the effect of typical magnetostrophic fields in three models: $fz$  fast-rotating ZAMS, $iz$ intermediate rotating ZAMS, $im$ intermediate rotating middle-aged MS stars. Vertical dashed lines intersecting the bottom x-axis correspond to the amount of differential rotation required to produce the same shift in the bi-layer, non magnetic model. \textit{Extracted from \citetalias{Barrault2025ExploringField}}. \textbf{Right}: Sample of $\gamma$ Dor stars containing dips analyzed by \citet{Saio2021RotationModes}. The x-axis denotes the detected differential rotation between the convective core and the near-core region. The y-axis is the central hydrogen fraction. Different data points for each star represent the results obtained for different prescriptions of core boundary mixing used. Two stars (KIC5985441 and KIC4390625) appear as outliers of the overall distribution. Core field strengths are overlaid on the original figure, denoting the amount of core magnetism able to make the two stars join the bulk of detected core-to-envelope differential rotation. \textit{Adapted from \citet{Saio2021RotationModes}, courtesy MNRAS}. }
\label{fig:2}
\end{figure}

The key difference shared by both the magnetic model \citepalias{Barrault2025ExploringField} and the hydrodynamical, bi-layer rotating one \citepalias{Barrault2025ConstrainingDips} and the one solidly rotating \citetalias{Tokuno2022AsteroseismologyOscillations} is the frequency shift introduced between the core and the envelope modes.
In this interaction where the core mode interacts with a series of prograde dipole envelope modes, hydrodynamical Bryan solutions require that the spin parameter of the pure inertial mode $s_{\rm core}^{*} = \dfrac{2\Omega_{\rm core}}{\sigma_{\rm core}}\simeq 11.32$, with $\sigma_{\rm core}$ the angular wave frequency in the frame co-rotating with the core. With the field configuration considered, we can show that this condition evolved to $\nu_{\rm M, core}^{*} = \dfrac{2(\Omega_{\rm core}\sigma_{\rm core}-m\omega_{\rm A,core}^{2})}{\sigma_{\rm core}^{2}-m^{2}\omega_{\rm A,core}^{2}}\simeq 11.32$. The spin parameter of the core mode is thus shifted with magnetism. The control parameter of this shift is shown to be the Lehnert number $\rm Le_{core} = \omega_{\rm A, core}/2\Omega_{\rm core}$, quantifying the relative effects of the Lorentz force and the Coriolis acceleration. Envelope modes in the TARM also undergo a magnetic shift \citep{Mathis2011Low-frequencyField}.\\
The result of the analysis on a typical inertial dip is shown on the right panel of Fig.\ref{fig:1}: for an increasing core's magnetic field from the fixed value of the envelope's magnetic field, the inertial dip is shifted towards low periods (or low $s_{\rm env}$, the spin parameter in the frame co-rotating with the envelope), and is getting thinner and deeper. Qualitatively, this can be understood as a magnetic equivalent of a Doppler shift. From the frame co-rotating with the core, the density of envelope gravito-inertial modes is decreased with increasing core magnetism, and thus the core $m\text{-}i$ modes will find a varying number of envelope gravito-inertial modes sufficiently close in frequency for a significant interaction to occur. The coupling efficiency is thus decreasing with an increasing magnetic contrast between the core and the envelope, as it is also the case considering only core-to-envelope differential rotation. Magnetism also makes the angular structure of both the core inertial modes and the envelope gravito-inertial modes change, which makes the mode geometrical matching vary at the core boundary. This effect is however not sufficiently grasped in the present model.

\subsection{Degeneracy with differential rotation}
\citetalias{Barrault2025ExploringField} shows that the signature of differential rotation and bi-layer toroidal magnetic fields are fully degenerated: for a set of rotations rates contrasts and near-core $N$, there exist a set of magnetic contrasts and near-core $N$ that creates a dip of same location and shape \citepalias{Barrault2025ExploringField}. This is no surprise, since under this configuration of fields the Lorentz force holds the same analytical form as the Coriolis acceleration. It remains to see whether this degeneracy can be lifted, either by an evaluation of the prevalence of both effects in typical $\gamma$ Dor stars or by the addition of some new constraints.\\
First, \citetalias{Barrault2025ExploringField} considers magnetic intensities arising from a typical dynamo in a magnetostrophic regime expected in the case of fast-rotating cores, for three realistic $\gamma$ Dor models providing an influence of age and rotation rates. The study shows that for fast-rotating ZAMS $\gamma$ Dors ($\Omega/2\pi \approx 2.5 \rm \, c.d^{-1}$), the effect of a typical magnetic field would be equivalent to the one of a differential rotation of $1.5\%$ between the core and the envelope. For an aged $\gamma$ Dor with intermediate rotation ($\Omega/2\pi \approx 1.5\, \rm \,c.d^{-1}$), the effect of typical magnetostrophic core fields would be the same as $5\%$ of differential rotation (see left panel of Fig.\ref{fig:2}). The magnetic signature would be washed out in the one of differential rotation in the first case, since a differential rotation of $1\%$ was shown to be reachable in realistic MHD simulations \citep{Featherstone2009EffectsStars}. On the contrary, we expect the magnetic field effect to be more prominent in the latter model, since a magnetic field of about $1 \, \rm MG$ potentially present in the core could significantly decrease core-to-envelope differential rotation.\\
However, a partial degeneracy still exists between the effects of differential rotation and core magnetism on the inertial dip. We expect that this can be lifted by a potential study of inertial dips present in the period spacing pattern of prograde modes of different azimuthal order $m$, since the differential rotation effect depends on $m$ and the magnetic effect on $m^2$. This is however a long-term goal, since no detection of inertial dip in prograde quadrupole mode series has ever been made up to now, and such a study will likely require a quality of data achieved only with the advent of the PLATO mission.\\
Since core magnetism and differential rotation have the same effect on the inertial dip's shift, we foresee that interesting targets for the detection of core magnetism will be the ones that show a dip strongly shifted towards low values of $s_{\rm env}$. Interestingly, two targets from the \citet{Saio2021RotationModes} stand out as having this property compared to a population of stars at the same core hydrogen content $X_{c}$. We propose for these two stars core field strengths able to create such a shift of the dip to join the remainder of the distribution (see Fig.\ref{fig:2}, right panel). We however point out that this cannot be considered as a detection and is only presented as a crude application of the present model.


\begin{acknowledgements}
L. Barrault and L. Bugnet gratefully acknowledge support from the European Research Council (ERC) under the Horizon Europe programme (Calcifer; Starting Grant agreement N$^\circ$101165631). S. Mathis acknowledges support from the PLATO CNES grant at CEA/DAp. S. Mathis and J.S.G. Mombarg acknowledge support from the European Research Council through HORIZON ERC SyG Grant 4D-STAR 101071505. While partially funded by the European Union, views and opinions expressed
are however those of the authors only and do not necessarily reflect those of the European Union or the European Research Council. Neither the European Union nor the granting authority can be held responsible for them. 
\end{acknowledgements}

\bibliographystyle{aa}  
\bibliography{Barrault_S20} 

\end{document}